%% file: MBradac.tex
\documentclass{evn2002}
\usepackage{graphicx}
\usepackage[]{natbib}
\bibpunct{(}{)}{;}{a}{}{,}
\begin{document}
\title{Using VLBI Data to Investigate the Galaxy Structure in the 
Gravitationally Lensed System B1422+231}
\titlerunning{Using VLBI Data to Investigate the Galaxy Structure in the 
GL System B1422+231}

   \author{M. Brada\v{c} \inst{1}
               \and
          P. Schneider \inst{1}
		\and
	  M. Steinmetz \inst{2}
		\and
	  M. Lombardi \inst{1}
		\and
          L. J. King \inst{1} 
		\and
	  R. W. Porcas \inst{3}	
          }

   \institute{Institut f\"{u}r Astrophysik und Extraterrestrische 
              Forschung, Auf dem H\"ugel 71, D-53121 Bonn, Germany
    \and Steward Observatory, 933 North Cherry Avenue, Tucson, AZ
   85721, USA
    \and Max-Planck-Institut f\"{u}r Radioastronomie, Auf dem
   H\"{u}gel 69, D-53121 Bonn, Germany
}		

%
%

\abstract{Gravitationally lensed systems with multiply imaged quasars are an
excellent tool for studying the properties of distant galaxies. In
particular, they provide the most accurate mass measures for the
lensing galaxy. The system B1422+231 is a well
studied example of a quadruply imaged quasar, with high-quality VLBI
data available. Very accurate data on image positions, fluxes and
deconvolved image sizes 
provide good constraints for lensing models. We discuss here the 
failure of smooth models in fitting the data. Since 
the mass of a lens galaxy is not a smooth entity, we have
investigated how deviation from a smooth model can influence lensing 
phenomena, especially the image flux ratios.
To explore expectations about the level of substructure in galaxies
and its influence on strong lensing, N-body simulations of a 
model galaxy are employed. By using the mass distribution of this
model galaxy as a lens, 
synthetic data sets of different four image system configurations are
generated. Their analysis can possibly provide
evidence for the presence and strong influence of substructure in the 
primary lens galaxy. 
}

\maketitle
%
%

\def\g{\gamma_1}
\def\gg{\gamma_2}
\def\eck#1{\left\lbrack #1 \right\rbrack}
\def\eckk#1{\bigl[ #1 \bigr]}
\def\rund#1{\left( #1 \right)}
\def\abs#1{\left\vert #1 \right\vert}
\def\wave#1{\left\lbrace #1 \right\rbrace}
\def\ave#1{\left\langle #1 \right\rangle}

\def\vc#1{%
  \if\alpha#1\mathchoice
    {\mbox{\boldmath$\displaystyle#1$}}%
    {\mbox{\boldmath$\textstyle#1$}}%
    {\mbox{\boldmath$\scriptstyle#1$}}%
    {\mbox{\boldmath$\scriptscriptstyle#1$}}%
  \else
    \textbf{\textit{#1}}%
  \fi}
\newcommand{\bams}[0]{Bull. Amer. Math. Soc}
%
%
\section{\label{sc:1422}The mystery of B1422+231}

The gravitational lens system B1422+231 was discovered in the course of 
the JVAS survey (Jodrell Bank -- VLA Astrometric Survey) by
\citet{pa92}. It consists of four image components. The three brightest 
images A, B, and C (as designated by \citealt{pa92}) are fairly 
collinear. The radio flux ratio between images A and B is approximately
0.9, while image C is fainter (flux ratio C to B is approximately
0.5). Image D is further away and is much fainter than the other images (with
flux ratio D:B of 0.03). 
We used
the most recent available radio data for the 
image positions and
fluxes from
the polarisation observations made at $8.4 \; {\rm GHz}$ using the
VLBA and the 100m telescope at Effelsberg (Patnaik et al.\ 1999).
For each of the components, the authors measured positions
(relative to the image B) and fluxes as well as the deconvolved image
shapes.

The radio source of this lens system is 
associated with a $15.5 \; {\rm mag}$ quasar at a redshift of $3.62$ 
\citep{pa92}. The 
lensing galaxy has been observed in the optical; its redshift has been 
determined 
to be $0.338$ and its position relative to image B has been measured 
\citep{im96}. The main lens 
galaxy is a member of a compact group with a median
projected radius of $35 \; h^{-1} \; {\rm kpc}$ and velocity dispersion 
of $\sim 550 \; {\rm km \: s^{-1} }$ \citep{ku97}.

Several groups have tried to model B1422+231
\citep{ho94,ko94b,ke97,ma98} and all of them have experienced
difficulties in fitting the image parameters. 
As we used data with even more precise image
positions one might expect that it would become even harder to model
the system. However, as already pointed out by some authors, the difficulties 
do not lie in fitting the image positions but rather in the
flux ratios.  All the results can be found in detail in \citet{br02}

\section{Lens modelling with smooth models\label{sc:lm}}
First we considered two standard gravitational lens models. We used a 
singular isothermal ellipsoid with
external shear from
\citet{ko94b} (hereafter SIE+SH) and a non-singular isothermal ellipsoid
model with external shear (NIE+SH) from \citet{ke98} 
to fit the image positions and fluxes of B1422+231.

We have applied the standard $\chi^2$ minimisation procedure 
to the radio data, using image positions, fluxes, and their uncertainties from
\citet{pa99}.    
The optical position of the galaxy was taken from
\citet{im96}. Although the image positions are very accurate 
(uncertainties of the order of $50 \;
{\mu \rm arcsec}$), we have no difficulties fitting them. However, 
as already pointed out in previous works on B1422+231,
such models completely fail
in predicting the image fluxes. In particular
image A is predicted too dim (the flux ratio A:B as predicted by
SIE+SH model is 0.80, much below the measured value of 0.93). 
We have also tried to model the system with a NIE+SH model; however, the
$\chi^2$ did not improve significantly.

Although other smooth models can still be investigated, it seems unlikely that
another smooth model can explain all four image fluxes
simultaneously. Even when one disregards the flux of image A, a smooth
macro-model seems to be incapable of explaining the remaining flux ratios.

\section{Models with substructure \label{sc:mss}}
The A:B flux ratio causes 
the biggest difficulty in 
fitting B1422+231. Since the radio and optical flux ratios are very
different, one is tempted to exclude it from the $\chi^2$ measure.

However, one can also try to deal with this problem in another
way. Adding a small perturber at the same angular diameter distance as
the primary lens and at approximately the same position as image A can
change the flux ratio A:B substantially. On the
other hand, calculations show \citep{ma98} that such a perturber does
not affect the positions of any of the images
appreciably. Furthermore, a small
perturber can also change the flux ratio of the other images slightly and
this might help to improve the results from the previous section.

We model the perturber as a non-singular isothermal sphere. 
We are aware of the fact that the choice of this particular model
for the substructure is oversimplified in many ways. However, we are
not trying to constrain the nature of substructure in this case, which
is impossible due to the number of constraints
available.

The resulting model has 12 parameters,
which leaves us 0 degrees of freedom. 
For a model with zero degrees of freedom we expect
$\chi^2$ to vanish if the model is realistic. The resulting $\chi^2 =
5.6$ remained high;
this family of models considered does not seem to be
adequate for the description of the galaxy in B1422+231 lens system.

\section{Using image shapes as constraints \label{sc:ell}}
For the more sophisticated models it is very
difficult to  ensure a constrained model that accounts for the
substructure using as constraints only image
positions, flux ratios, and the galaxy position.  
For this reason we also included  the axis ratios 
and orientation angles of the deconvolved images from \citet{pa99}
as additional constraints.

It turned out that it is difficult to simultaneously stretch and rotate
the images with one (or two) perturber(s). The macro-model is not very
successful in predicting both components of the image ellipticities
and the fluxes, and
therefore corrections are needed in the case of all four images. 
We can safely assume that the inclusion of further sub-clumps in
the model would eventually lead to a ``perfect'' fit with the observed
data. In particular three more sub-clumps close to the images B, C, and
D would yield a significant improvement to the $\chi^2$ and could
explain the observed data. 

\section{Strong lensing by an N-body simulated galaxy \label{sc:nbody}}
A question that arises from (difficulties in) model fitting of B1422+231 is 
whether such behaviour is seen with an N-body simulated galaxy and
therefore generic of a typical galaxy lens.
We used the cosmological N-body simulation data including 
gas dynamics and star formation of \citet{st01} for this purpose.

\begin{figure}
\begin{minipage}{0.33\textwidth}
\includegraphics[width=5cm]{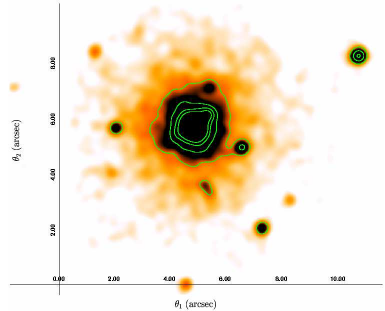}
\end{minipage}
\begin{minipage}{0.15\textwidth}
\caption{The cut-out of the surface mass density map of the simulated 
galaxy. Note that the scale of this figure is wrong in \citet{br02}.}
\end{minipage}
         \label{fig:kappa_2k}
   \end{figure}

The lens properties are calculated using the IMCAT software. 
We use this ``lens'' 
to generate systems that have 
similar configuration and flux ratios as in
the case of B1422+231. In total we considered 11 different
synthetic systems. 

The fitting procedure is performed in the same way as for the
B1422+231 data. Again, image B is taken as a reference. We
try to fit the positions and fluxes with SIE+SH and SIE+SH+NIS
models.  We experience similar problems 
fitting fluxes as before; the $\chi^2$-function value is high for 
all 11 data sets. 

There are indications that the level of substructure as obtained from
simulations can
influence lensing phenomena a lot. In particular, the synthetic fluxes we
obtained deviate strongly from those predicted by smooth models. 
This particular example of a simulated galaxy can of course not
give us the answers to the aforementioned questions. 
To draw stronger conclusions, one would have to investigate many
different realisations of N-body simulated galaxies and in addition
use higher resolution simulations (currently unavailable). A
statistical analysis to investigate the strong lensing properties 
could then be made. 

\begin{acknowledgements}
M.B.\ acknowledges partial support from the EC ICN
RadioNET (Contract No. HPRI-CT-1999-40003).
\end{acknowledgements}
\vspace{-5pt}
\bibliography{bibliogr} 
\bibliographystyle{aa}
\end{document}

%% file: MBradac.bbl
\begin{thebibliography}{11}
\expandafter\ifx\csname natexlab\endcsname\relax\def\natexlab#1{#1}\fi

\bibitem[{{Brada\v{c}} {et~al.}(2002){Brada\v{c}}, {Schneider}, {Steinmetz},
  {Lombardi}, {King}, \& {Porcas}}]{br02}
{Brada\v{c}}, M., {Schneider}, P., {Steinmetz}, M., {et~al.} 2002, \aap, 388,
  373

\bibitem[{{Hogg} \& {Blandford}(1994)}]{ho94}
{Hogg}, D. \& {Blandford}, R. 1994, \mnras, 268, 889

\bibitem[{{Impey} {et~al.}(1996){Impey}, {Foltz}, {Petry}, {Browne}, \&
  {Patnaik}}]{im96}
{Impey}, C., {Foltz}, C., {Petry}, C., {Browne}, I., \& {Patnaik}, A. 1996,
  \apjl, 462, L53

\bibitem[{{Keeton} \& {Kochanek}(1998)}]{ke98}
{Keeton}, C. \& {Kochanek}, C. 1998, \apj, 495, 157

\bibitem[{{Keeton} {et~al.}(1997){Keeton}, {Kochanek}, \& {Seljak}}]{ke97}
{Keeton}, C., {Kochanek}, C., \& {Seljak}, U. 1997, \apj, 482, 604

\bibitem[{{Kormann} {et~al.}(1994){Kormann}, {Schneider}, \&
  {Bartelmann}}]{ko94b}
{Kormann}, R., {Schneider}, P., \& {Bartelmann}, M. 1994, \aap, 286, 357

\bibitem[{{Kundic} {et~al.}(1997){Kundic}, {Hogg}, {Blandford}, {Cohen},
  {Lubin}, \& {Larkin}}]{ku97}
{Kundic}, T., {Hogg}, D., {Blandford}, R., {et~al.} 1997, \aj, 114, 2276

\bibitem[{{Mao} \& {Schneider}(1998)}]{ma98}
{Mao}, S. \& {Schneider}, P. 1998, \mnras, 295, 587

\bibitem[{{Patnaik} {et~al.}(1992){Patnaik}, {Browne}, {Walsh}, {Chaffee}, \&
  {Foltz}}]{pa92}
{Patnaik}, A., {Browne}, I., {Walsh}, D., {Chaffee}, F., \& {Foltz}, C. 1992,
  \mnras, 259, 1P

\bibitem[{{Patnaik} {et~al.}(1999){Patnaik}, {Kemball}, {Porcas}, \&
  {Garrett}}]{pa99}
{Patnaik}, A., {Kemball}, A., {Porcas}, R., \& {Garrett}, M. 1999, \mnras, 307,
  L1

\bibitem[{{Steinmetz} \& {Navarro}(2001)}]{st01}
{Steinmetz}, M. \& {Navarro}, J. 2001, {NewA submitted}

\end{thebibliography}
